\documentclass[prd,showpacsfloatfix,superscriptaddress,nofootinbib]{revtex4}
\usepackage{amsmath,amssymb,epsfig,bm}
\newcommand{\be}{\begin{equation}}
\newcommand{\ee}{\end{equation}}
\newcommand{\bea}{\begin{eqnarray}}
\newcommand{\eea}{\end{eqnarray}}

\newcommand{\vecp}{\bm p}
\newcommand{\vecQ}{\bm Q}
\newcommand{\veck}{\bm k}

\def\ie{{\it i.e.}}
\def\eg{{\it e.g.}\,, }

\begin{document}
\title{Phase diagram of chiral quark matter:
color and electrically neutral Fulde-Ferrell phase}
\date{June 1, 2010}
\author{Xu-Guang Huang}
\affiliation{Institute for Theoretical Physics, J. W. Goethe-University
D-60438 Frankfurt am Main, Germany} \affiliation{Frankfurt Institute
for Advanced Studies, D-60438 Frankfurt am Main, Germany }

\author{Armen Sedrakian}
\affiliation{Institute for Theoretical Physics, J. W. Goethe-University
D-60438 Frankfurt am Main, Germany}

\begin{abstract}
The phase diagram of charge and color neutral two-flavor color 
superconducting quark matter is studied including the homogeneous 
two-flavor superconductor (2SC) and the inhomogeneous Fulde-Ferrell 
(FF) phases within the Nambu-Jona-Lasinio model. The low-temperature 
domain $T\le 5$~MeV of the phase diagram contains the FF phase, 
which borders at high temperatures to the 2SC phase. The critical temperature
of phase transition from the 2SC to the unpaired state is in the range
20-30 MeV. We derive the equation of state of matter 
and its composition and show that matter in mature compact stars should be 
in the inhomogeneous FF-like superconducting state. We briefly discuss the
astrophysical implications of such a phase in compact stars. 
\end{abstract}

\maketitle

\section{Introduction}
\label{sec:introduction}

The quark substructure of baryons, predicted by quantum
chromodynamics (QCD), suggests that nucleonic matter may
deconfine at sufficiently high densities attainable in compact
stars. Compact stars featuring quark cores, known as hybrid
stars, may be distinguishable from their purely nucleonic/hyperonic
counterparts by their astrophysical manifestations, which may
include magnetic, thermal, and rotational evolutions on secular
timescales, as well as shorter transients.

In this paper we derive the equation of state and composition
of two-flavor dense quark matter at nonzero temperature under the
conditions of charge neutrality and $\beta$ equilibrium.
At moderate densities quark matter is composed of up and down
quark flavors only, which in newborn protoneutron stars occur at
near equal amount (the isospin chemical is vanishingly small).
However, rapid compression of matter leads to its neutronization 
via the inverse beta decay and the isospin chemical potential grows
until it reaches the asymptotic value determined by 
the $\beta$-equilibrium condition
$\mu_d-\mu_u =\mu_e$ among $d$ and $u$ quarks and electrons
(here $\mu_i$ with $i=d,\, u,\, e$ are their chemical potentials,
respectively). Therefore, under stellar conditions the pairing between the two
light flavors of quarks occurs at finite isospin chemical potential,
\ie, when the Fermi surfaces of up and down quarks are shifted apart
by an amount which could be of the same order of magnitude as the gap
in the quasiparticle spectrum. Under these conditions
the actual pairing pattern may be significantly different from
that of the BCS and is likely to involve breaking of the spatial
symmetries by the condensate order parameter, in analogy to
less exotic low-temperature systems found in condensed matter
(\eg electron gas or ultracold atomic
vapor)~\cite{Fulde:1965,Larkin:1965,Takada:1969}.

The phase diagram of charge and color neutral quark matter, which features
{\it homogeneous} condensates at non-zero temperature and moderate densities,
is well established at the mean-field 
level~\cite{Ruester:2005jc,Blaschke:2005uj,Warringa:2005jh}.
It contains a sequence of phases with the two-flavor superconducting (2SC) 
phase at low densities and the color-flavor-locked (CFL) phase at 
asymptotically high densities (the color superconducting 
phases are reviewed in Ref.~\cite{Alford:2007xm}; 
for a review of astrophysical implications 
of color superconductivity see, \eg Ref.~\cite{Alford:2009qj,Sedrakian:2009uu}).
The domain occupied by the {\it inhomogeneous} phases is difficult to access
in general, since most of the studies are carried out in the regime
where the Ginzburg-Landau expansion of the mean-field thermodynamic
potential is valid~\cite{Bowers:2002xr,Casalbuoni:2005zp,Rajagopal:2006ig,
Mannarelli:2006fy,Ippolito:2007uz}. 
An exception is the so-called Fulde-Ferrell 
(or plane-wave crystalline color superconducting)
phase~\cite{Alford:2000ze,Giannakis:2005,Fukushima:2006su,Kiriyama:2006ui,He:2006vr,Sedrakian:2009kb,Andersen:2010vu}, which admits straightforward
treatment in the full domain of interest to compact stars. 
Low-dimensional treatments of inhomogeneous phases admit some 
simplifications as well~\cite{Nickel:2008ng}. In 
effective four-fermion coupling models, like the Nambu-Jona-Lasinio 
(NJL) model, the interactions can be easily varied in a broad 
range of (effective) couplings that may account for medium modifications 
of the interactions in dense quark matter.

In this paper we examine the phase diagram of charge and color neutral
quark matter at non-zero temperatures and moderate densities and show
that the simplest realization of the phases with broken spatial symmetries
-- the Fulde-Ferrell phase -- arises in the low-temperature
domain of the phase diagram. We do so, by extending the discussion in
Ref.~\cite{Sedrakian:2009kb} to include 
the conditions of charge and color neutrality,
which impose additional constraints on the set of equations that 
are solved.
As indicated above, we have chosen the spatial dependence of the 
order parameter that corresponds to a single plane wave.
Therefore, our calculations offer only an upper bound on the
ground state energy, since more complicated order 
parameters have been proven
to possess lower energy in the domain where the Ginzburg-Landau expansion 
is valid. However, qualitatively the phases with broken spatial symmetries 
share many common features and the Fulde-Ferrell (FF) 
phase is a convenient starting point of exploration.

Our discussion is restricted to the two-flavor case appropriate at
moderate densities, \ie, its validity domain is constrained by the
temperature and density where strange quarks appear in substantial
amounts. The role of strange quarks is twofold: first, because of
their negative charge it becomes energetically favorable to
replace high momentum electrons by low momentum strange quarks; 
second, strange quarks disfavor the BCS type cross-flavor 
pairing among the light quarks and favors instead the FF or
other crystalline-type pairing. 
If the Fermi energy of strange quarks is large
enough they may choose to pair in a color-flavor-locked pattern which,
except for the asymptotically high densities, will involve some mismatch
between the Fermi surfaces with possibly multiple FF phases involving
quarks of all three flavors.

Finally, a number of non-crystalline phases have been suggested in the 
literature, which address the stress caused by the mismatch of Fermi surfaces 
differently. These alternative phases invoke separation into normal
and superconducting domains~\cite{Bedaque:2003hi} or pairing induced 
changes in the shapes of the Fermi surfaces~\cite{Muther:2002mc}. 

This paper is organized as follows. Section~\ref{sec:key_equations}
presents the key equations describing the FF phase (see also
Ref.~\cite{Sedrakian:2009kb}). In Sec.~\ref{sec:neutrality}
we discuss the conditions of charge and color neutrality. Our results
are presented in Sec.~\ref{sec:results} and our conclusions are
collected in Sec.~\ref{sec:summary}.

\section{Gap equations and thermodynamics of FF phase}
\label{sec:key_equations}

The mean field (MF) 
thermodynamic potential of the FF phase is given by
\bea \Omega_{\rm MF} = \Omega^{\Delta}_{\rm MF} + {\Omega}^{0}_{\rm MF} ,
\eea where
the first term includes the contribution from the condensate of
red-green quarks and the second term is the contribution from the
unpaired blue quarks. The first term is given by
\bea\label{Omega_Delta} 
{\Omega}^{\Delta}_{\rm MF}&=&3\frac{\Lambda^2\Delta^2}{g^2}
-2\sum_{e}\int\frac{d^3k}{(2\pi)^3}\Bigl\{
E_e^{+}(\Delta)+E_e^{-}(\Delta) -2T~{\rm ln}\,
f\left[-E_e^{+}(\Delta)\right] -2T~{\rm ln}\,
f\left[-E_e^{-}(\Delta)\right] \Bigr\}, 
\eea 
where $\Lambda$ is the cutoff of the NJL model, $g$ is the 
strong coupling constant, $f(\omega)
=\left[1+\exp(\omega/T)\right]^{-1}$ is the Fermi
distribution, and $T$ is the temperature.
The contribution of the blue quarks is
\bea {\Omega}^{0}_{\rm MF} &=&-
2\int\frac{d^3k}{(2\pi)^3}\sum_f\left\{|{\bf k}|-T{\rm
ln}f\left[-\beta(|\veck|-\mu_b^f)\right]-T{\rm
ln}f\left[-\beta(|\veck|+\mu_b^f)\right]\right\}, \eea
where $f$ summation is over flavors and $\mu_b^f$ is the chemical 
potential of blue quarks of flavor $f$.
The thermodynamic potential depends on the 
quasiparticle energies, which are given by
\bea
E_e^{\pm}(\Delta)&\approx& E_{A,e}\pm \sqrt{E_{S,e}^{2}
+\vert\Delta\vert^2},
\eea
where $E_{S,e}$ and $E_{A,e}$ are the parts of the spectrum
which are even (symmetric) and odd (asymmetric) under exchange
of quark flavors in a Cooper pair. These are given by\footnote{Because 
$\vert\vecQ\vert \sim \delta\mu$ is small compared with
the typical scale $\bar\mu$ for quark momenta, we can keep the
leading order in $\vert\vecQ\vert/\bar\mu$ terms, in which
case the spectrum takes the form given in Ref.~\cite{Sedrakian:2009kb}.}

\bea
\label{spectrum_S}
E_{S,e}(\vert\veck\vert,\vert\vecQ\vert,\theta,\bar\mu) &=&E^+ -e\bar\mu,\\
\label{spectrum_A}
E_{A,e}(\vert\veck\vert,\vert\vecQ\vert,\theta,\delta\mu) &=&
\delta\mu+eE^-,
\eea
with $E^{\pm} = (\vert \veck+\vecQ/2\vert \pm
\vert \veck-\vecQ/2\vert)/2$, where $\theta$ is
the angle formed by vectors $\veck$ and $\vecQ$, $\delta\mu =
(\mu_u-\mu_d)/2$ and $\bar\mu = (\mu_u+\mu_d)/2$.
The spectrum of unpaired blue quarks is given by
$\xi_{e}^{b}(\veck,\mu_b)=E_{S,e}(\veck,0,0,\mu_b)=\vert\veck\vert-
e\mu_b$, where $\mu_b$ is the chemical potential of blue
quarks.

The form of quasiparticle spectrum requires a subtraction from the
thermodynamic potential of the unphysical contribution from currents
that exist in the case where pairing gap vanishes. Therefore, we need
to evaluate
\be \Omega_{\rm MF} = {\Omega}_{\rm MF}-{\Omega}_{\rm
MF}^{\Delta=0}+{\Omega}_{\rm MF}^{\Delta=0,{\bf Q}=0}.
\ee
The stationary point(s) of the thermodynamic
potential determine the equilibrium values of the order parameters:
\be\label{eq:derivatives} 
\frac{\partial\Omega_{\rm
MF}}{\partial\Delta} = 0, \quad \frac{\partial\Omega_{\rm
MF}}{\partial\vert \vecQ\vert} = 0.
\ee
The direction of the vector $\vecQ$, \ie, the axis of symmetry breaking
is chosen by the system spontaneously; the appearance of a preferred direction 
implies that $O(3)$ symmetry of the BCS theory is broken down to $O(2)$ in
the FF phases; phases with more complicated lattice-type order parameters 
break the translational symmetry of the problem. 
The relative concentrations of the $u$ and $d$ quarks are
determined from the derivative of the thermodynamic
potential with respect to the corresponding chemical potential as
\be \frac{\partial\Omega_{\rm MF}}{\partial\mu_u} = n_u, \quad
\frac{\partial\Omega_{\rm MF}}{\partial\mu_d} = n_d. \ee

The explicit form of the gap equation, which follows 
from Eqs. (\ref{Omega_Delta}) and (\ref{eq:derivatives}), is
\bea\label{eq:gap_final} \Delta
&=&\frac{g^2}{3\Lambda^2 } \sum_e\int\frac{d^3p}{(2\pi)^3}
\frac{\Delta}{E_e^{+}(\Delta)-E_e^{-}(\Delta)} \left\{ {\rm
tanh}\left[\frac{\beta}{2}E_e^{+}(\Delta)\right]
-{\rm tanh}\left[\frac{\beta}{2}E_e^{-}(\Delta)\right] \right\}.
\eea 
Note that in the isospin-symmetric limit the quasiparticle spectra 
are degenerate $E_e^{+}(\Delta)=-E_e^{-}(\Delta)$ and 
(with $\vert\vecQ\vert =0$) the gap equation reduces to the 
ordinary BCS gap equation for ultrarelativistic fermions.
Finally, the equation for the total momentum is given by
\bea\label{eq:tot_momentum} 0 &=&
-\sum_{e}\int\frac{d^3k}{(2\pi)^3}
\left\{\frac{\partial E_e^+(\Delta)}{\partial
|\vecQ|}\tanh\left[\frac{E_e^{+}(\Delta)}{2T}\right]
+\frac{\partial E_e^-(\Delta)}{\partial
|\vecQ|}\tanh\left[\frac{E_e^{-}(\Delta)}{2T}\right]
\right\},
\eea
with
\bea
\frac{\partial E_e^{\pm}(\Delta)}{\partial
|\vecQ|}=\left[\frac{e}{2}
\pm\frac{E_{S,e}(|\veck|,|\vecQ|,\theta,\bar\mu)}{E_e^+(\Delta)-E_e^-(\Delta)}\right]
\frac{Q+2k\cos\theta}{2|2\veck+\vecQ|}-
\left[\frac{e}{2}\mp\frac{E_{S,e}(|\veck|,|\vecQ|,\theta,\bar\mu)}{E_e^+(\Delta)-E_e^-(\Delta)}\right]
\frac{Q-2k\cos\theta}{2|2\veck-\vecQ|}.
\eea
The zero temperature counterparts of 
Eqs.~(\ref{eq:gap_final})-(\ref{eq:tot_momentum}) are
straightforward to obtain with the help of the identity ${\rm
tanh}(x/2) = 1-2f(x)$ and the limiting form of the Fermi
distribution function $f(x) = \theta (-x)$ for $T\to 0$,
where $\theta(x)$ is the Heaviside step function.

As long as strange quarks are too heavy to appear in matter, the 
net positive charge of quarks is neutralized by electrons. At relevant 
densities and temperatures the electrons can be treated to a good approximation 
as a noninteracting gas. Their thermodynamic potential is then given by 
\bea \label{Omega_el}
\Omega_{\rm elec}=2T\int\frac{d^3k}{(2\pi)^3}\left[
{\rm ln}f\left(-\xi_{\rm elec}^-(\veck,\mu_e)\right)
+{\rm ln}f\left(-\xi_{\rm elec}^+(\veck,\mu_e)\right)
\right],
\eea
with $\xi_{\rm elec}^-(\veck,\mu_e)=\sqrt{k^2+m_e^2}-\mu_e$, 
where $m_e$ is the electron mass. The second term in Eq.~(\ref{Omega_el})
represents the contribution of anti-particles (positrons) to the 
thermodynamical potential with $\xi_{\rm elec}^+(\veck,\mu_e)
=\sqrt{k^2+m_e^2}+\mu_e$.

\section{Imposing charge and color neutrality}
\label{sec:neutrality}

The matrix representing the chemical potentials of quarks 
in the color-flavor space can be written in terms of the color 
group generators and the matrix of electrical charge operators as
\bea\label{mu_matrix}
\mu=\frac{\mu_B}{3}-\mu_eQ_e+\mu_3 T^c_3+\mu_8 T^c_8, \eea
where $\mu_B$ is the baryon chemical potential,
the electrical charge operator in flavor space is
\bea \label{Qoperator}
Q_e={\rm
diag}\left(\frac{2}{3},-\frac{1}{3},-\frac{1}{3}\right),
\eea
while the color operators are given by
\bea\label{Toperator}
T^c_3=\frac{1}{2}{\rm diag}(1,-1,0), 
\quad T^c_8=\frac{1}{2\sqrt{3}}{\rm diag}(1,1,-2).
\eea
We choose the direction of the color symmetry breaking the blue direction. The
densities of red and green quarks are then automatically equal, therefore
$n_3\equiv0$ and there is no need for the conjugate to $n_3$ chemical
potential $\mu_3$. From Eqs. (\ref{mu_matrix})-(\ref{Toperator}),
\bea\label{MU1}
\bar{\mu} &=& \frac{\mu_B}{3}-\frac{\mu_e}{6}+\frac{\mu_8}{2\sqrt{3}},\\
\label{MU2}
\delta\mu &=& -\frac{\mu_e}{2}.
\eea
The full thermodynamic potential of our system is now given by
$\Omega=\Omega_{\rm MF}+\Omega_{\rm elec}$. By choosing instead of chemical 
potentials of quarks the chemical potentials $\mu_8$ and $\mu_e$ 
as independent thermodynamic variables we obtain a set
of equations that should be solved simultaneously:
\bea\label{GAP3}
\frac{\partial\Omega_{\rm MF}}{\partial\Delta} = 0, \quad
\frac{\partial\Omega_{\rm MF}}{\partial\vert \vecQ\vert} = 0,
\eea
\bea \label{GAP4}
\frac{\partial\Omega}{\partial\mu_e} = 0, \quad
\frac{\partial\Omega}{\partial\mu_8} = 0. \eea
The densities of quarks and leptons can be derived through derivatives
of the thermodynamic potential with respect to the conjugate chemical
potentials; for electron density we have
\bea n_{\rm elec} = -\frac{\partial\Omega_{\rm elec}}{\partial\mu_e}
&=&2\int\frac{d^3k}{(2\pi)^3}\left[f(\xi_{\rm elec}^+)-f(\xi_{\rm
elec}^-)\right].
\eea
The densities of paired up and down quarks are given by
\bea n_d^{rg}+n_u^{rg}
&=&-\frac{\partial\Omega}{\partial\bar\mu}=
-\sum_{e}\int\frac{d^3k}{(2\pi)^3}\frac{eE_{S,e}}{\sqrt{E_{S,e}^{2}
+\Delta^2}}\left[ \tanh\left(\frac{E_e^{+}(\Delta)}{2T}\right)-
\tanh\left(\frac{E_e^{-}(\Delta)}{2T}\right)\right]
,\\
n_u^{rg}-n_d^{rg} &=&-\frac{\partial\Omega}{\partial\delta\mu}=
\sum_{e}\int\frac{d^3k}{(2\pi)^3}\left[
\tanh\left(\frac{E_e^{+}(\Delta)}{2T}\right)+
\tanh\left(\frac{E_e^{-}(\Delta)}{2T}\right)\right].
\eea
The densities of unpaired up and down quarks are given by
\bea n_d^{b} &=&
2\sum_{e}\int\frac{d^3k}{(2\pi)^3}ef(\xi_{e}(\veck,\mu^b_d))
,\\
n_u^{b} &=&
2\sum_{e}\int\frac{d^3k}{(2\pi)^3}ef(\xi_{e}(\veck,\mu^b_u)). \eea
The total densities of
up and down quarks are
\bea n_d&=&n_d^{rg}+n_d^b,\\
n_u&=&n_u^{rg}+n_u^b.
\eea
The baryon density is
\bea n_B &=&\frac{n_u+n_d}{3}.
\eea
The electrical and charge neutrality require, respectively,
\bea
\frac{2}{3}n_u-\frac{1}{3}n_d-n_{\rm elec} &=& 0,\\
n_d^{rg}+n_u^{rg}-2n_d^b-2n_u^b &=& 0.
\eea
The thermodynamic properties of matter are completely determined through
four gap equations (\ref{GAP3})-(\ref{GAP4}) for a given temperature
and baryon density (or baryon chemical potential).

\section{Results}
\label{sec:results}
\begin{table}[htb]\renewcommand{\arraystretch}{1.5}\addtolength{\tabcolsep}{0.01pt}
\caption{
The column entries are as follows: baryonic density in units of fm$^{-3}$,
total pressure and electron pressure in units of MeV fm$^{-3}$, the
baryonic chemical potential, the chemical potentials of $u$, $d$ 
quarks and electrons 
in units of MeV, the densities of $d$ quarks, $u$ quarks, and electrons in
units of fm$^{-3}$, the gap in units of MeV, 
and total momentum of the condensate in units of fm$^{-1}$.
}\label{table1}
\begin{tabular}{cccccccccccc} \hline\hline
$n_B\quad$ & $P_{\rm tot}\quad$ & $P_{\rm e}\quad$ & $\mu_B\quad$ & $\mu_d\quad$ & $\mu_u\quad$ & $\mu_e\quad$ & $n_d\quad$ & $n_u\quad$ & $n_e\quad$ & $\Delta\quad$ & $|\vecQ|$ \\
\hline
\multicolumn{12}{c}{$T=1$ {MeV}}\\
\hline
0.32 & 63.9 & 27.9 & 1316.0 & 498.7 & 474.1 & 24.6 & 0.43 & 0.55 & 0.14 & 38.0 & 0.21 \\

0.36 & 76.5 & 32.6 & 1370.6 & 519.6 & 494.0 & 25.6 & 0.49 & 0.61 & 0.16 & 45.8 & 0.29 \\

0.40 & 90.2 & 37.6 & 1422.0 & 539.4 & 512.9 & 26.5 & 0.55 & 0.68 & 0.18 & 53.7 & 0.37 \\ 
0.44 & 105.0 & 42.7 & 1470.8 & 558.2 & 530.9 & 27.4 & 0.62 & 0.75 & 0.20 & 61.9 & 0.45 \\ 
0.48 & 121.0 & 47.9 & 1517.2 & 576.3 & 548.1 & 28.2 & 0.68 & 0.81 & 0.21 & 70.2 & 0.53 \\ 
\hline
\multicolumn{12}{c}{$T=5$ {MeV}}\\
\hline
0.32 & 66.5 & 29.0 & 1322.5 & 501.5 & 475.8 & 25.7 & 0.44 & 0.55 & 0.14 & 37.8 & 0.00 \\

0.36 & 77.1 & 32.8 & 1369.4 & 519.2 & 493.3 & 25.9 & 0.49 & 0.61 & 0.16 & 45.7 & 0.00 \\

0.40 & 90.7 & 37.6 & 1420.1 & 538.6 & 512.0 & 26.6 & 0.55 & 0.67 & 0.18 & 53.7 & 0.00 \\ 
0.44 & 105.6 & 42.7 & 1468.4 & 557.2 & 529.7 & 27.5 & 0.61 & 0.74 & 0.20 & 61.8 & 0.18 \\ 
0.48 & 121.7 & 48.0 & 1514.7 & 575.1 & 546.8 & 28.3 & 0.67 & 0.80 & 0.21 & 70.2 & 0.29 \\ 
\hline
\multicolumn{12}{c}{$T=10$ {MeV}}\\
\hline
0.32 & 62.7 & 26.7 & 1309.5 & 495.6 & 474.6 & 21.0 & 0.43 & 0.53 & 0.14 & 35.9 & 0.00 \\
0.36 & 86.6 & 39.3 & 1370.8 & 523.1 & 488.0 & 35.2 & 0.46 & 0.62 & 0.16 & 43.6 & 0.00 \\
0.40 & 100.4 & 44.1 & 1422.8 & 542.5 & 507.7 & 34.8 & 0.53 & 0.69 & 0.18 & 52.7 & 0.00 \\
0.44 & 111.3 & 46.3 & 1468.9 & 558.8 & 527.6 & 31.1 & 0.60 & 0.74 & 0.19 & 61.4 & 0.00 \\ 
0.48 & 126.1 & 50.5 & 1513.9 & 575.6 & 545.2 & 30.4 & 0.66 & 0.80 & 0.21 & 70.0 & 0.00 \\ 
\hline
\hline
\end{tabular}\label{table}
\end{table}
\begin{figure}[t]
\vskip 0.9cm
\begin{center}
\includegraphics[height=10.0cm,width=10.5cm,angle=0]{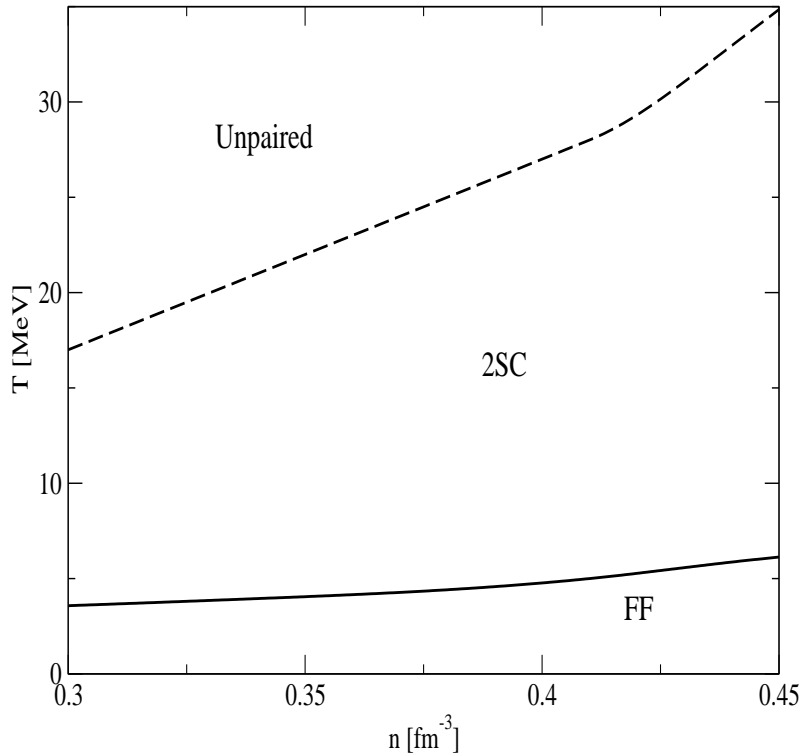}
\end{center}
\caption[]
{ (color online).
The phase diagram of neutral two-flavor quark matter. The high-temperature
low-density region corresponds to unpaired quark matter, the intermediate
temperature region corresponds to the homogeneous 2SC phase, and the
low-temperature domain corresponds to the inhomogeneous FF phase
of quark matter.
}\label{fig:phase_diagram_1}
\end{figure}

Eqs.~(\ref{GAP3}) and (\ref{GAP4}) were solved numerically by finding a solution
that satisfies simultaneously the four integral equations. The ultraviolet
divergence of the integrals was regularized by a three-dimensional cut-off
in momentum space $\vert\vecp\vert < \Lambda$. The phenomenological
value of the coupling constant
$G = g^2/12\Lambda^2$ in the $\langle qq\rangle$
Cooper channel is related to the coupling constant $G_d$
in the $\langle\bar
qq\rangle$ di-quark channel by the relation $G=N_c/(2N_c-2)G_d$,
where $N_c=3$ is the number of quark colors. The coupling constant
$G_d$ and the cut-off $\Lambda$ are fixed by adjusting the model
to the vacuum mass and decay constant of the pion.
As in Ref.~\cite{Sedrakian:2009kb} we employ the parameter
values $G_d=3.11$ GeV$^{-2}$ and $G_d\Lambda^2 = 1.31$.

The phase diagram of charge and color neutral matter is shown in 
Fig.~\ref{fig:phase_diagram_1}.
The baryonic density range extends up 
to $3\rho_0$, where $\rho_0 = 0.16$ fm$^{-3}$ 
is the nuclear saturation density. The model 
is limited to the above density range, 
because at somewhat larger densities the Fermi 
momentum becomes comparable to the 
cut-off of the model. The covered temperature range extends up to the critical 
temperature for the onset of color superconducting phase transition. It is seen 
that the low-temperature domain is occupied by the FF phase; in 
the intermediate temperature 
regime $5\le T \le 20$ MeV, 
the FF phase is replaced by a homogeneous 2SC phase, the
phase transition being first order. 
At the critical temperature, which depending 
on the density lies in the range $T_c\sim 20-30$ MeV, 
there is a second order phase 
transition from the 2SC to the unpaired phase. 

The equation of state of matter and related parameters are tabulated in
Table~\ref{table1}. The large pairing gaps at large densities are due 
to the density scaling of the density of states at the Fermi surface; 
note that the magnitude of the gap depends on 
the chosen value of the cutoff and is, therefore, specific to the present 
model. Furthermore, the pressure requires some normalization procedure, which 
is chosen to be the standard NJL prescription, which requires zero 
pressure (and energy 
density) in vacuum (\ie, at $T=\mu=0$). In practice, the low density limit 
of the model should correspond to quarks confined in baryons, so that the 
normalization should be controlled by the confinement mechanism in non-zero 
temperature/density QCD. This freedom
allows one to add a bag constant to the pressure of the NJL model to match 
it to the equation of state of hadronic matter at a desired density 
and temperature of deconfinement. 
The point of onset of deconfinement is, thus, a free parameter 
of the model. Consequently, the pressure of the model may be shifted by a constant value of the bag.

As seen from Table~\ref{table1}, 
the equation of state is only weakly temperature dependent. 
Table~\ref{table1} shows also the composition of matter 
and the chemical potentials of species, which are related by the 
$\beta$-equilibrium condition $\mu_d=\mu_e+\mu_u$. This condition will
be violated in the high temperature domain $T\ge 10$ MeV of our phase 
diagram, where neutrinos are thermally populated 
as a consequence of their short mean-free path at such temperatures. 
The non-zero neutrino chemical potential then must be added to the 
left-hand  side of the $\beta$-equilibrium condition (\ref{MU2}).
Since there is no or little overlap between the domain occupied by the FF 
phase and the domain where neutrinos are trapped, we shall not repeat our
analysis at non-zero neutrino chemical potential. Let us note that 
the phase transition from the 2SC to the FF phase is first order and is 
associated with a latent heat; such a heat release in compact stars during
the collapse stage may serve as an engine that powers
high-energy burst phenomena from the collapsing star.

\section{Summary}
\label{sec:summary}

In this paper we have investigated the 
phase diagram of charge and color neutral 
quark matter in two-flavor NJL model. Our results strongly 
suggest that the 2SC phase does not exist in the interiors of mature 
compact stars. Instead, the stellar matter at intermediate densities 
is in the FF, or related crystalline, phase. A possible 
alternative, which is not excluded by our calculations, 
is that nuclear matter directly transforms to three-flavor 
CFL matter. The FF phase provides an upper bound on the 
energy of the color superconducting phase, which can presumably be
lowered by choosing more complicated forms of the order parameter. 
The model parameters in our study were chosen to correspond to the 
common values adopted in the NJL model; in the unlikely 
case where the in-medium couplings are 
substantially larger, the 2SC phase may have lower 
energy than the FF phase (see Ref.~\cite{Sedrakian:2009kb}).

Provided a nuclear equation of state and a matching 
procedure (\eg Maxwell construction) the present 
equation of state can be used to construct the complete 
equation of state of matter in compact stars and their 
models (such constructions are reviewed in, \eg
Refs.~\cite{WEBER_BOOK,Sedrakian:2006mq}). 
The interesting possibility that three-flavor crystalline phase may appear 
in stable massive compact stars has been pointed out 
recently~\cite{Ippolito:2007hn}.
The presence of a crystallinelike phase in the interiors of 
hybrid compact stars may have a number of 
interesting astrophysical implications; \eg 
the finite shear modulus of matter implies deformations of the quark core that 
can generate gravitational waves at the level of the sensitivity of current 
gravitational wave observatories~\cite{Knippel:2009st}. 
The properties of the FF phase may manifest themselves in the 
cooling of compact stars. The emission rate of the dominant 
Urca process has been computed in phases similar to the FF phase
in Refs.~\cite{Huang:2007cr,Anglani:2006br}.
Finally, the first order BCS-FF phase 
transition and the latent heat release associated with it
may power high-energy burst phenomena from collapsing stars.

\acknowledgements
We thank D. H. Rischke for discussions.
This work was supported, in part, by the Alliance Program 
of the Helmholtz Association, contract HA216/EMMI {\it ``Extremes of 
Density and Temperature: Cosmic Matter in the Laboratory''}, the 
Helmholtz International Center for FAIR within the framework of the LOEWE
program of the State of Hesse and by the Deutsche
Forschungsgemeinschaft (Grant SE 1836/1-1).

\end{document}